\documentclass[a4paper,onecolumn,11pt]{article}
\usepackage{amsfonts}
\usepackage{graphicx}
\usepackage{amsmath}

\input{tcilatex}
\begin{document}

\title{A Blade Tip-Timing method based on Periodic Nonuniform Sampling of
order 2 }
\author{B. Lacaze \\
%EndAName
TeSA Laboratory, 7 Bd de la Gare, 31500 Toulouse France\\
\textit{e-mail: }bernard.lacaze@tesa.prd.fr}
\maketitle

\begin{abstract}
Vibrations are among main causes of fatigue and damages leading to
destruction of rotating blades. Consequently, motions of blades have to be
carefully studied, and in particular, periodic components. Blade Tip-Timing
(BTT) methods achieve non-intrusive measurements which is a great advantage
with respect to classical methods. They can be modelled by a Periodic
Nonuniform Sampling of order $L$ equal to the number of probes (PNS$L$). In
this paper, we develop the case $L$=2, and we prove that this model is
sufficient for finding the places of asynchronous frequency lines, with
their amplitudes and their phases. To increase $L$ will increase the
redundance and then the accuracy of the method.

\textit{keywords: }blade tip-timing, periodic nonuniform sampling,
undersampling.
\end{abstract}

\section{Introduction}

Blade Tip-Timing (BTT) addresses a device including

\ \ \ \ \ \ \ \ \ \ \ - one or several blades rotating at constant velocity

\ \ \ \ \ \ \ \ \ \ \ - one or several probes which measure the transit time
of the endpoints of blades at each turn \cite{Heat}, \cite{Zheng}.

Knowing the velocity of blades, the device measures deviations with respect
to an ideal behavior. For a given blade, results are modelled by a
stationary real random process $\mathbf{X}=\left\{ X\left( t\right) ,t\in 
\mathbb{R}\right\} ,$ sampled at times%
\begin{equation}
t_{kn}=nT+\theta _{k},1\leq n\leq N,1\leq k\leq L
\end{equation}%
where $L$ is the number of probes and $N$ the number of turns taken into
account. The $\theta _{k}$ define the relative positions of the probes. The
value of $N$ is large enough to consider first the set%
\begin{equation}
t_{kn}=nT+\theta _{k},n\in \mathbb{Z},1\leq k\leq L.
\end{equation}%
We recognize the sampling plan PNS$L$ (Periodic Nonuniform Sampling of order 
$L)$ \cite{Yen}, \cite{Lin}$.$ For well-chosen $\theta _{k},$ this sampling
plan is able to overpass $\left( L-1\right) $ spectral foldings. The
"period" $T$ is dictated by the geometry of the studied material (the
dimension of blades and the rotation speed). For an errorless reconstruction
of the process \textbf{X }in the frequency band $\left( -a,a\right) ,$ the
Nyquist condition in the PNS$L$ case limits the value of $a$ to \cite{Papo}%
\begin{equation*}
a_{\max }=\frac{L}{2T}.
\end{equation*}%
Therefore, to increase the number of probes $L$ allows to increase the
length of the frequency band allowed by the Nyquist condition. However, this
condition is not optimal when the spectral support of \textbf{X} does not
contain neighbourhoods of the frequency 0. In this situation, the Landau
condition is better, using only the length of the spectral support rather
than its location \cite{Land}. An undersampled plan in the Nyquist sense is
often adequate in the Landau sense. Two-bands processes used in
communications provide a good example. To use the Landau condition in the
BTT method does not solve problems of errorless reconstruction, but it
allows to consider multiple sampling plans in the same time. It will
increase the total of available information.

In this paper, we shows that the PNS$2$ plan is sufficient to retrieve
together the place, the amplitude and the phase of asynchronous frequency
lines \cite{Laca1}, \cite{Laca2}. Section 2 explains this property, using a
matched reconstruction formula (proved in Appendix). Section 3 gives a plan
for retrieving spectral lines when their number is not too large.

\section{A formula for PNS2}

1) Let assume that $\mathbf{X}=\left\{ X\left( t\right) ,t\in \mathbb{R}%
\right\} $ is a real zero-mean stationary process with power spectrum $%
s\left( f\right) $ defined as%
\begin{equation*}
E\left[ X\left( t\right) X\left( t-\tau \right) \right] =\int_{\Delta
_{k}}e^{2i\pi f\tau }s\left( f\right) df
\end{equation*}%
\begin{equation}
\Delta _{k}=\left( -\frac{k+1}{T},\frac{-k}{T}\right) \cup \left( \frac{k}{T}%
,\frac{k+1}{T}\right)
\end{equation}%
where E$\left[ ..\right] $ stands for the mathematical expectation and where 
$k$ is some positive or nil integer$.$ $\Delta _{k}^{+}$ and $\Delta
_{k}^{-} $ are the positive and negative parts of $\Delta _{k}.$ Here, $%
s\left( f\right) $ will be an even real function, addition of a regular
function and a finite sum of "Dirac functions" \cite{Papo}. Actually, the
last component models frequency lines, which are accumulations of power at
the neighbourhood of some places. The following formula is proved in
appendix (when $\alpha \theta /T\notin \mathbb{Z}$)%
\begin{equation}
\left\{ 
\begin{array}{l}
X\left( t\right) =\frac{-A\left( t\right) \sin \left[ \pi \alpha \left(
t-\theta \right) /T\right] +B\left( t\right) \sin \left[ \pi \alpha t/T%
\right] }{\sin \left[ \pi \alpha \theta /T\right] },\alpha =2k+1 \\ 
A\left( t\right) =\sum_{n\in \mathbb{Z}}\left( -1\right) ^{n}\text{sinc}\pi
\left( \frac{t}{T}-n\right) X\left( nT\right) \\ 
B\left( t\right) =\sum_{n\in \mathbb{Z}}\left( -1\right) ^{n}\text{sinc}\pi
\left( \frac{t-\theta }{T}-n\right) X\left( nT+\theta \right) .%
\end{array}%
\right.
\end{equation}%
From any process \textbf{Y}, we define the process $\widetilde{\mathbf{Y}}$
by 
\begin{equation}
\left\{ 
\begin{array}{l}
\widetilde{Y}\left( t\right) =\frac{-A\left( t\right) \sin \left[ \pi \alpha
\left( t-\theta \right) /T\right] +B\left( t\right) \sin \left[ \pi \alpha
t/T\right] }{\sin \left[ \pi \alpha \theta /T\right] },\alpha =2k+1 \\ 
A\left( t\right) =\sum_{n\in \mathbb{Z}}\left( -1\right) ^{n}\text{sinc}\pi
\left( \frac{t}{T}-n\right) Y\left( nT\right) \\ 
B\left( t\right) =\sum_{n\in \mathbb{Z}}\left( -1\right) ^{n}\text{sinc}\pi
\left( \frac{t-\theta }{T}-n\right) Y\left( nT+\theta \right) .%
\end{array}%
\right.
\end{equation}%
We have $\widetilde{\mathbf{Y}}=\mathbf{Y}$ when conditions $\left( 3\right) 
$ are fulfilled for $\mathbf{Y}$.\ 

2) Formula $\left( 4\right) $ above is true for $Y\left( t\right) =e^{2i\pi
f_{0}t}$ when $f_{0}$ is inside $\Delta _{k}.$ Now, let assume that%
\begin{equation}
\left\{ 
\begin{array}{l}
f_{0}=\overline{f_{0}}+\underline{f_{0}} \\ 
\overline{f_{0}}\in \Delta _{k}^{+},\underline{f_{0}}\in \frac{1}{T}\mathbb{Z%
}\text{.}%
\end{array}%
\right.
\end{equation}%
This decomposition of $f_{0}$ is unique (except perhaps at the bounds), and
figure 1 gives an illustration. Formulas $\left( 4\right) $ are true for $%
Y\left( t\right) =e^{2i\pi \overline{f_{0}}t}$. If we address these formulas
to $Y\left( t\right) =e^{2i\pi f_{0}t}$ with $f_{0}$ $\notin \Delta _{k},$
we obtain a result $\widetilde{e^{2i\pi f_{0}t}}$ different from $e^{2i\pi
f_{0}t}:$ 
\begin{equation}
\widetilde{e^{2i\pi f_{0}t}}=\frac{e^{i\pi \underline{f_{0}}\theta }}{\sin 
\frac{\pi \alpha \theta }{T}}\left[ e^{2i\pi \overline{f_{0}}t}\sin \pi
\theta \left( \frac{\alpha }{T}+\underline{f_{0}}\right) -e^{2i\pi t(%
\overline{f_{0}}-\frac{\alpha }{T})+i\pi \frac{\alpha \theta }{T}}\sin \pi
\theta \underline{f_{0}}\right] .
\end{equation}%
Therefore, the spectral line at $f_{0}$ is broken down in two lines, the
first one at $\overline{f_{0}}\in \Delta _{k}^{+}$ and the second one at $%
\overline{f_{0}}-\frac{\alpha }{T}\in \Delta _{k}^{-}$ (both quantities do
not depend on $\theta ).$ The same work can be done for $e^{-2i\pi f_{0}t}.$
By linearity, we obtain the behavior $\widetilde{\cos 2\pi f_{0}t}$ of a
"real spectral line" $Y\left( t\right) =\cos 2\pi f_{0}t,f_{0}$ as $\left(
6\right) ,$ when introduced in $\left( 5\right) :$%
\begin{equation}
\widetilde{\cos 2\pi f_{0}t}=\frac{1}{\sin \frac{\pi \alpha \theta }{T}}%
\left[ \cos \left( 2\pi \overline{f_{0}}t+\pi \underline{f_{0}}\theta
\right) \sin \pi \theta \left( \frac{\alpha }{T}+\underline{f_{0}}\right)
\right.
\end{equation}%
\begin{equation*}
\left. -\cos \left( 2\pi t\left( \frac{\alpha }{T}-\overline{f_{0}}\right)
-\pi \theta (\underline{f_{0}}+\frac{\alpha }{T})\right) \sin \pi \theta 
\underline{f_{0}}\right] .
\end{equation*}%
We verify that, for $\underline{f_{0}}=0$ (i.e $f_{0}\in \Delta _{k}^{+}),$
or $\underline{f_{0}}=-\left( 2k+1\right) /T$ (i.e $f_{0}\in \Delta
_{k}^{-}),$\ we retrieve 
\begin{equation*}
\widetilde{\cos 2\pi f_{0}t}=\cos 2\pi f_{0}t.
\end{equation*}%
In both cases, $f_{0}\in \Delta _{k},$ and $\left( 4\right) $ is available.

3) \textbf{Consequences}

The introduction in formula $\left( 5\right) $ of the "real spectral line" $%
Y\left( t\right) =\cos 2\pi f_{0}t,$ $f_{0}\notin \Delta _{k},$ highlights
two terms on the positive axis of frequencies:

a) a line at $\overline{f_{0}}\in \Delta _{k}^{+}$ with amplitude $\sin \pi
\theta \left( \frac{\alpha }{T}+\underline{f_{0}}\right) /\sin \frac{\pi
\alpha \theta }{T}$ and phase $\underline{f_{0}}\theta /2\overline{f_{0}}$
(the first term in $\left( 8\right) ),$

b) a line at $\frac{\alpha }{T}-\overline{f_{0}}\in \Delta _{k}^{+}$ with
amplitude $\sin \pi \underline{f_{0}}\theta /\sin \frac{\pi \alpha \theta }{T%
}$ and phase $\theta (\underline{f_{0}}+\frac{\alpha }{T})/2\left( \overline{%
f_{0}}-\frac{\alpha }{T}\right) $ (the last term in $\left( 8\right) ).$ The
frequency $\frac{\alpha }{T}-\overline{f_{0}}$ is the symmetric of $%
\overline{f_{0}}$ with respect of $\frac{\alpha }{2T}$ (the middle of $%
\Delta _{k}^{+}).$

Any real line in $\Delta _{j},j\neq k,$ is folded in two parts on $\Delta
_{k}^{+},$ with a weight depending only on $j,k,$ and with a position
independent of $\theta .$ It is possible that one of both lines disappears
(when its weight cancels). It is the case when $\underline{f_{0}}\theta $ or 
$\theta \left( \frac{\alpha }{T}+\underline{f_{0}}\right) $ is an integer.

Let consider some process \textbf{Y, }a sum of a "noise" \textbf{N} and a
collection of spectral lines $\lambda _{1},\lambda _{2}...$ Samples $Y\left(
nT\right) ,Y\left( nT+\theta \right) $ are plugged in $\left( 5\right) ,$
for some $k$ ($\alpha =2k+1).$ The noise \textbf{N }becomes a new noise $%
\widetilde{\mathbf{N}},$ result of foldings of \textbf{N }in $\Delta _{k}.$
The lines $\lambda _{j}$ which belong to $\Delta _{k}$ are not changed. The
lines $\lambda _{j}$ outside $\Delta _{k}$ are folded in two parts on $%
\Delta _{k},$ with amplitude and phase functions well identified of $\theta $
(the relative positions of both probes), of $T$ (the geometry and the
celerity of the material in study), and of $\overline{f_{0}},\underline{f_{0}%
}$ (relative positions of $f_{0}$ with respect to $\Delta _{k}).$ The
spectral study of the result for several values of $k,$ is sufficient to
estimate the characteristics of the lines $\lambda _{j},$ assuming that the
foldings can be distinguished.

A synchronous spectral line corresponds to a frequency close to $f=k^{\prime
}/T$ for some integer $k^{\prime }.$ If we assume that it occupates a narrow
interval $\left( \frac{k^{\prime }}{T}-\varepsilon ,\frac{k^{\prime }}{T}%
+\varepsilon \right) ,$ this line gives two repliquas close to $\frac{k}{T}$
and $\frac{k+1}{T}.$ Then, when using $\left( 5\right) ,$ synchronous lines
are viewed close to the bounds $\frac{k}{T}$ and $\frac{k+1}{T}$ and
asynchronous lines occupate the inside of the interval $\Delta _{k}^{+}.$

\section{Diagnosis about asynchronous spectral lines}

We note $M_{k}$ the number of spectral lines in $\Delta _{k}^{+}$\ when
using $\left( 5\right) $ with samples of a process \textbf{Y} which possibly
does not fulfill condition $\left( 3\right) $%
\begin{equation*}
s\left( f\right) =0,f\notin \Delta _{k}.
\end{equation*}%
We assume that all foldings of spectral lines are distinct, and particularly
that $f_{j}\notin \frac{1}{2T}+\frac{1}{T}\mathbb{Z}$. We assume that $%
\underline{f_{0}}\theta $ and $\theta \left( \frac{\alpha }{T}+\underline{%
f_{0}}\right) $ are not integers, so that folded lines at $\underline{f_{0}}$
and at $\frac{\alpha }{T}-\underline{f_{0}}$ do not cancel out. We neglect
synchronous spectral lines which appear around the points $\frac{k}{T},\frac{%
k+1}{T}.$

Let assume that $k=0.$ We have the following situations:

\begin{quotation}
\textbf{M}$_{0}=0:$
\end{quotation}

No spectral line.

\begin{quotation}
\textbf{M}$_{0}=1:$
\end{quotation}

1 line, unchanged by the algorithm.

\begin{quotation}
\textbf{M}$_{0}=2:$
\end{quotation}

1 line outside $\Delta _{0}.$ We have $\overline{f_{1}}+\overline{f_{2}}$ $%
=1/T,$ and we look for \underline{$f_{1}$}$=$\underline{$f_{2}$}. It
suffices to find the $\Delta _{k}$ such that $M_{k}=1.$

\begin{quotation}
\textbf{M}$_{0}=3:$
\end{quotation}

a) 3 lines in $\Delta _{0}$ when $M_{k}=6,k\neq 0$

b) 1 line in $M_{0},$ and 1 outside. The last one verifies $\overline{f_{2}}+%
\overline{f_{3}}$ $=2/T.$ We have $M_{k}=3$ or 4, and this line is inside
the $\Delta _{k},k\neq 0,$ such that $M_{k}=3$.

\begin{quotation}
\textbf{M}$_{0}=4:$
\end{quotation}

a) 4 lines in $\Delta _{0}$ when $M_{k}=8,k\neq 0.$

b) 2 lines in $\Delta _{0}$ and 1 line outside when $M_{k}=5$ or 6$,k\neq 0.$
When $M_{k}=5,$ we have found the line outside $\Delta _{0}.$

c) 0 line in $\Delta _{0},$ and 2 outside. When $M_{k}=2$ or 3, we have
found the place of the lines.

\begin{quotation}
\textbf{M}$_{0}=5:$
\end{quotation}

a) 5 lines in $\Delta _{0}$ when $M_{k}=10,k\neq 0.$\ 

b) 3 lines in $\Delta _{0}$ and 1 line outside when $M_{k}=7$ or 8, $k\neq 0$%
. The first value provides the line outside $\Delta _{0}$

c) 1 line in $\Delta _{0}$ and 2 lines outside when $M_{k}=4,5$ or 6, $k\neq
0$. The first value when we reach a $\Delta _{k}$ with 2 lines, the second
with one line.

\begin{quotation}
\textbf{M}$_{0}=6:$
\end{quotation}

a) 6 lines in $\Delta _{0}$ when $M_{k}=12,k\neq 0.$

b$)$ 4 lines in $\Delta _{0}$ and 1 outside, when $M_{k}=9$ or 10, $k\neq 0$%
. The first value provides the line outside.

c) 2 lines in $\Delta _{0}$ and 2 outside, when $M_{k}=6,7$ or 8, $k\neq 0.$
The first value when we reach a $\Delta _{k}$ with 2 lines, the second with
one line.

d) 0 lines in $\Delta _{0}$ and 3 outside, when $M_{k}=3,5$ or 6, $k\neq 0.$
The first value when we reach a $\Delta _{k}$ with 3 lines, the second with
2 lines, the third with one line...

The same routine can be made from other values of $k.$ When we deal with a
baseband process, we will use $k=0,$ but, for a two-bands process (as in
communications), $k$ will be chosen so that $\Delta _{k}$ belongs to the
used frequencies.

To illustrate the procedure above, we consider data taken in \cite{Bouc} for
simulations. Authors consider a device rotating at the frequency $1/T=470$%
Hz. The number of registered turns is $L=150$. Five probes are located by
angles 3.6, 33.6, 144, 291.6, 313.2 in degrees. A mixing of two asynchronous
signals of frequencies $f_{1}=213$Hz and $f_{2}=1131$Hz is studied. The
respective amplitudes are 10 and 1 with phases 0.3 and 0.5 radians. The
noise is not very well defined.

Whatever the two chosen probes, we have%
\begin{equation*}
M_{0}=3,M_{1}=4,M_{2}=3,M_{k}=4,k\geq 3.
\end{equation*}%
Conversely, this result implies the existence of one and only one line in $%
\Delta _{0}$ and $\Delta _{2},$ where only one line is doubled. Figure 2
shows the locations of lines. The height of lines depends on the relative
position of chosen probes. The choice of different combinations of probes
will allow to obtain a sufficient redundancy for improving accuracy.

\section{The noise}

The "noise" $\mathbf{N}=\left\{ N\left( t\right) ,t\in \mathbb{R}\right\} $%
\textbf{\ }is the component of \textbf{X} with a continuous power spectrum,
which cumulates with spectral lines. We have%
\begin{equation*}
N\left( t\right) =\sum_{n=0}^{\infty }N_{n}\left( t\right)
\end{equation*}%
where the process $\mathbf{N}_{n}$ is the part of $\mathbf{N}$ within the
frequency band $\Delta _{n}.$ These components are uncorrelated. When $%
\alpha =2k+1,$ when $\mathbf{N}$ is the input of $\left( 5\right) ,$ and $%
\widetilde{\mathbf{N}}$ is the output, we obtain from $\left( 8\right) $%
\begin{equation*}
\widetilde{N}\left( t\right) =\sum_{j=0}^{\infty }\frac{N_{j}\left( t\right) 
}{\sin \left( 2k+1\right) \frac{\pi \theta }{T}}\left[ \cos \left[ \frac{%
2\pi }{T}\left( j-k\right) \left( t-\frac{\theta }{2}\right) \right] \sin
\left( j+k+1\right) \frac{\pi \theta }{T}\right.
\end{equation*}%
\begin{equation}
\left. -\cos \left[ \frac{2\pi }{T}\left( j+k+1\right) \left( t-\frac{\theta 
}{2}\right) \right] \sin \left( j-k\right) \frac{\pi \theta }{T}\right] .
\end{equation}%
As explained in section 5 below, each term of the sum $\left( 9\right) $ is
non-stationary (except for $j=k)$. As usual, the noise will limit the
efficiency of the method. It may be possible to improve the signal on noise
ratio using a prefilter cancelling too high frequencies, hoping that
asynchronous lines are not zaped.

\section{Remarks}

1) In practice, devices provide an accurate knowledge of parameters $T$
(through the rotation celerity), $\theta \left( T\right) $ (through the
place of probes). $\alpha =2k+1$ being chosen, formula $\left( 5\right) $ is
applied from data. We obtain a process with power spectrum in $\Delta _{k}$.
More precisely, we have estimations because the number of data is limited
(two times the number of turns)$.$ We deduce a set of spectral estimations
for a large enough number of values of $k$ (depending on the number of
probes)$,$ which indicate the place of asynchronous lines. Theoretically,
the amplitude and the phase can be estimated.

The estimation of power spectra addresses old and well documented
techniques. For instance, the Prony method goes back to 1795 \cite{Haue}. A
large review of methods of spectral analysis is given in \cite{Stoi}, \cite%
{Babu}. A recent method is explained in \cite{Laca3}, \cite{Bona1}.

2) If it exists one synchronous wave, it will appear in each $\Delta _{k}$,
at each bound $\frac{k}{T}$ and $\frac{k+1}{T}.$ If they are several, they
cumulate. The method does not give the number and the places (the harmonic
number). As $T$ is known, crosscorrelations are likely to give the right
tool to solve this problem. Existence of synchronous lines is not
surprising, and it can produce numerous harmonics. To have the fundamental
and its harmonics confused at the bounds of $\Delta _{k}$ is a good
property, which forbids multiplication of synchronous lines mixed with
asynchronous one.

3) The spectral line notion is an idealization. Actually, it is an
accumulation of power on thin frequencial intervals, but with a
non-negligible width. Moreover, computations weaken contrasts. This explains
that the right place of lines can be not accurately defined, and that it
will be difficult to discriminate between lines that are too close to each
other. Nevertheless, estimations can be improved, because informations for a
given line are obtained from several frequency bands $\Delta _{k}.$

4) The research of the spectral lines is based on the knowledge of $M_{k},$
for several values of $k.$ Difficulties will come from the noise level,
which is folded, and then will increase with $T.$ Places of lines are
independent from $\theta $, but folded lines are weighted by $\sin \pi 
\underline{f_{0}}\theta $ and $\sin \pi \theta \left( \frac{\alpha }{T}+%
\underline{f_{0}}\right) $ ($\underline{f_{0}}$ depends on $\alpha ),$ and
these quantities can be very weak. However, while noted, the problem becomes
redundant, provided that a large enough number of $M_{k}$ estimations is
available.

5) The basic hypothesis is done in the section above: we assume that all
foldings of spectral lines are distinct (except at the bounds of $\Delta
_{k} $ where synchronous waves may appear). The hypothesis is verified if
and only if $f_{j}-f_{k}\notin \frac{1}{T}\mathbb{Z}$, whatever $j,k$ (for
ideal lines without width, which is an approximation). When the condition is
not verified for different indices $j_{0},k_{0},$ both lines will be folded
at the same place, wathever $\Delta _{k},$ $k\neq j_{0}$ and $k_{0}$.
However, it is different for $f_{j_{0}},f_{k_{0}}$ with respect to $\Delta
_{j_{0}}$ or $\Delta _{k_{0}}:$ $f_{j_{0}}$ will be retrieved in $\Delta
_{j_{0}}$ and $f_{k_{0}}$ will be split in two parts.

6) The algorithm confuses lines $f_{j}\in \frac{1}{2T}+\frac{1}{T}\mathbb{Z}%
, $ which are not split in two parts $\left( \overline{f}_{j}=\frac{k}{T}-%
\overline{f}_{j}\text{or }\frac{k+1}{T}-\overline{f}_{j}\right) .$ This case
can be linked to hidden synchronous lines, for instance via nonlinearities
in motorization. Conversely, a line at the middle of $\Delta _{k},$ is the
marker of this property. Section 3 is still available, disregarding what
happens in the middle of $\Delta _{k}^{+}$ and $\Delta _{k}^{-}.$

7) Let assume that the power spectrum of the stationary process \textbf{Y }%
cancels outside $\Delta _{j}^{-}$ for instance. For such a (complex)
process, $\left( 7\right) $ is equivalent to a sum of two amplitude
modulations, because \underline{$f_{0}$} is a constant when $f_{0}\in $ $%
\Delta _{j}^{-}.$ Both lead to a stationary result, in the form $\left(
f_{1}\neq f_{2}\right) $%
\begin{equation*}
Y_{1}\left( t\right) =\alpha Y\left( t\right) e^{2i\pi f_{1}t},Y_{2}\left(
t\right) =\beta Y\left( t\right) e^{2i\pi f_{2}t}
\end{equation*}%
but the sum is not stationary, because the cross-correlation between \textbf{%
Y}$_{1}$ and \textbf{Y}$_{2}$ depends on $t.$ We note that the time average
of the $\left( \mathbf{Y}_{1}+\mathbf{Y}_{2}\right) $-correlation function
eliminates $t.$

8) Formulas $\left( 7\right) $ and $\left( 8\right) $\ summarize motions and
changes in amplitude and phase of the spectral line $e^{2i\pi f_{0}t}$
induced by formula $\left( 5\right) .$ Actually, \ the phase of the input
does not matter much. If we consider the spectral line $e^{2i\pi f_{0}\left(
t+\phi \right) },$ we see that the foldings locations and amplitudes are
unchanged$.$ It is not the case for the phase.

\section{Conclusion}

1) The "sampling formula" (Shannon, Nyquist, Kotelnikov...) 
\begin{equation}
U\left( t\right) =\sum_{n=-\infty }^{\infty }\text{sinc}\pi \left( \frac{t}{T%
}-n\right) U\left( nT\right)
\end{equation}%
is true for a process \textbf{U }with power spectrum cancelling for
frequencies outside $\left( -1/2T,1/2T\right) $ \cite{Papo}$.$ When this
property fails, foldings will appear. For instance, let define \textbf{V }by%
\begin{equation*}
V\left( t\right) =U\left( t\right) +Ae^{2i\pi f_{0}t}
\end{equation*}%
where \textbf{U }verifies $\left( 10\right) $ and where $E\left[ A\right]
=0,E\left[ \left\vert A\right\vert ^{2}\right] <\infty $ (for staying in the
stationary framework, and with $A$ and $\mathbf{U}$\textbf{\ }
uncorrelated). If $\left\vert f_{0}\right\vert <1/2T,$ formula $\left(
10\right) $ is true for \textbf{V} 
\begin{equation*}
V\left( t\right) =\sum_{n=-\infty }^{\infty }\text{sinc}\pi \left( \frac{t}{T%
}-n\right) V\left( nT\right)
\end{equation*}%
If $\left\vert f_{0}\right\vert >1/2T,$ we have%
\begin{equation*}
U\left( t\right) +Ae^{2i\pi f_{1}t}=\sum_{n=-\infty }^{\infty }\text{sinc}%
\pi \left( \frac{t}{T}-n\right) V\left( nT\right)
\end{equation*}%
where $f_{1}=f_{0}-N/T,N\in \mathbb{Z},$ $f_{1}\in \left( -1/2T,1/2T\right)
. $ This means that formula $\left( 10\right) $ has folded the line from $%
f_{0} $ to $f_{1}$ by a multiple of $1/T.$ Conversely, if formula $\left(
10\right) $ shows a single line at $f_{1},$ its true place may be any $%
f_{0}=f_{1}+N/T$, except if we are sure that the spectrum cancels outside $%
\left( -1/2T,1/2T\right) .$ Without this information, we cannot decide what
is the true location of this line (i.e in what interval $\left( \left(
2N-1\right) /2T,\left( 2N+1\right) /2T\right) $ lies $f_{0})$.

This property takes another shape in the PNS2 context, where we consider
two-bands processes in $\Delta _{k},$ and samples at times $nT$ and $%
nT+\theta ,n\in \mathbb{Z}$. Formula $\left( 10\right) $ becomes formula $%
\left( 4\right) .$ Being given some line at $f_{0},$ this line will be
recovered if it belongs to $\Delta _{k}.$ Otherwise, the line will be
doubled and this property is generally sufficient to decide on the
membership of the line at any $\Delta _{k}$ and also to measure its
amplitude and its phase (varying the value of $k).$

2) Blade Tip-Timing (BTT) is a non-intrusive method for detecting spectral
lines in the blade motion, using particular sampling plans defined by
probes, each of them looking at deformations at periodic times. To overcome
undersamplings, the number of probes is increased, which adds coasts of
components, and leads to more complex computations. Actually, it is possible
to suppress these difficulties, holding the problem in the Landau context
rather than using Nyquist bounds, i.e working with unions of intervals
rather than with an unique interval centered at the origin. Using the above
property about foldings, we show that a device with only two probes is
theoretically sufficient to highlight asynchronous vibrations. It is
possible to elaborate PNS2 plans (Periodic Nonuniform Sampling of order 2)
for solving this problem, with a large choice of solutions.

The stationarity is the only hypothesis necessary to justify the results,
and the compactness of spectra is an useless hypothesis. We consider
processes with spectral lines which model undesirable vibrations. We show
that algorithms that are unsuitable for reconstruction yield characteristic
shiftings and/or splittings up which allow to retrieve the lines properties.
Lines are idealizations of power concentrations with a width which will be
increased \ by computations following experiments. Adequacy of theoretical
results with measurements (to be done) is a function of both the original
process (line purity) and algorithms.

This paper shows that true informations may be extracted from wrong
formulas. Provided that enough parameters have some freedom in these
formulas (in this case a band localization), it is possible to reach
expected estimations. Section 2 gives results obtained when misusing the
basic formula $\left( 4\right) $ fitted to PNS2. It is shown that an
asynchronous line will be viewed as unique or double, following its
location. Section 3 deduces a routine to locate asynchronous spectral lines,
based on variations of an unique parameter, the location of the used
frequency bands.

\section{Appendix}

Let assume that%
\begin{equation*}
s\left( f\right) =0,f\notin \Delta _{k}=\left( -\frac{k+1}{T},-\frac{k}{T}%
\right) \cup \left( \frac{k}{T},\frac{k+1}{T}\right)
\end{equation*}%
with $k\geq 0\mathbf{.}$ We have%
\begin{equation*}
X\left( t\right) =X^{+}\left( t\right) +X^{-}\left( t\right)
\end{equation*}%
where \textbf{X}$^{+}$ is the part of \textbf{X }in the frequency band $%
\Delta _{k}^{+}=\left( \frac{k}{T},\frac{k+1}{T}\right) $ and \textbf{X}$%
^{-} $ the part in $\Delta _{k}^{-}=\left( -\frac{k+1}{T},-\frac{k}{T}%
\right) .$ The Fourier series development of $e^{2i\pi ft}$ on $\Delta
_{k}^{+}=\left( \frac{k}{T},\frac{k+1}{T}\right) $ leads to both formulas
(for fixed $f$ different of $\frac{k}{T},\frac{k+1}{T},$ and with sinc$%
x=\left( \sin x\right) /x)$%
\begin{equation*}
e^{2i\pi ft}=\left\{ 
\begin{array}{c}
\sum_{n\in \mathbb{Z}}\left( -1\right) ^{n}e^{i\pi \alpha t/T}\text{sinc}\pi
\left( \frac{t}{T}-n\right) e^{2i\pi fnT} \\ 
\sum_{n\in \mathbb{Z}}\left( -1\right) ^{n}e^{i\pi \alpha (t-\theta )/T}%
\text{sinc}\pi \left( \frac{t-\theta }{T}-n\right) e^{2i\pi f(nT+\theta )}
\\ 
f\in \left( \frac{k}{T},\frac{k+1}{T}\right) ,\alpha =2k+1.%
\end{array}%
\right.
\end{equation*}%
Consequently (when $s\left( f\right) $ is regular at the bounds):%
\begin{equation*}
X^{+}\left( t\right) =\left\{ 
\begin{array}{c}
\sum_{n\in \mathbb{Z}}\left( -1\right) ^{n}e^{i\pi \alpha t/T}\text{sinc}\pi
\left( \frac{t}{T}-n\right) X^{+}\left( nT\right) \\ 
\sum_{n\in \mathbb{Z}}\left( -1\right) ^{n}e^{i\pi \alpha (t-\theta )/T}%
\text{sinc}\pi \left( \frac{t-\theta }{T}-n\right) X^{+}\left( nT+\theta
\right)%
\end{array}%
\right.
\end{equation*}%
and also%
\begin{equation*}
X^{-}\left( t\right) =\left\{ 
\begin{array}{c}
\sum_{n\in \mathbb{Z}}\left( -1\right) ^{n}e^{-i\pi \alpha t/T}\text{sinc}%
\pi \left( \frac{t}{T}-n\right) X^{-}\left( nT\right) \\ 
\sum_{n\in \mathbb{Z}}\left( -1\right) ^{n}e^{-i\pi \alpha (t-\theta )/T}%
\text{sinc}\pi \left( \frac{t-\theta }{T}-n\right) X^{-}\left( nT+\theta
\right) .%
\end{array}%
\right.
\end{equation*}%
We obtain both linear equations%
\begin{equation}
\left\{ 
\begin{array}{c}
X^{+}\left( t\right) e^{-i\pi \alpha t/T}+X^{-}\left( t\right) e^{i\pi
\alpha t/T}=A\left( t\right) ,\alpha =2k+1 \\ 
X^{+}\left( t\right) e^{-i\pi \alpha \left( t-\theta \right) /T}+X^{-}\left(
t\right) e^{i\pi \alpha \left( t-\theta \right) /T}=B\left( t\right) \\ 
A\left( t\right) =\sum_{n\in \mathbb{Z}}\left( -1\right) ^{n}\text{sinc}\pi
\left( \frac{t}{T}-n\right) X\left( nT\right) \\ 
B\left( t\right) =\sum_{n\in \mathbb{Z}}\left( -1\right) ^{n}\text{sinc}\pi
\left( \frac{t-\theta }{T}-n\right) X\left( nT+\theta \right)%
\end{array}%
\right.
\end{equation}%
which imply (when $\alpha \theta /T\notin \mathbb{Z}$) 
\begin{equation}
X\left( t\right) =\frac{-A\left( t\right) \sin \left[ \pi \alpha \left(
t-\theta \right) /T\right] +B\left( t\right) \sin \left[ \pi \alpha t/T%
\right] }{\sin \left[ \pi \alpha \theta /T\right] },\alpha =2k+1.  \label{4}
\end{equation}%
The method can also be applied to more general sets in the PNS$n$ framework 
\cite{Laca3}.

\begin{figure}[t]
\centering
\includegraphics[width = 14cm]{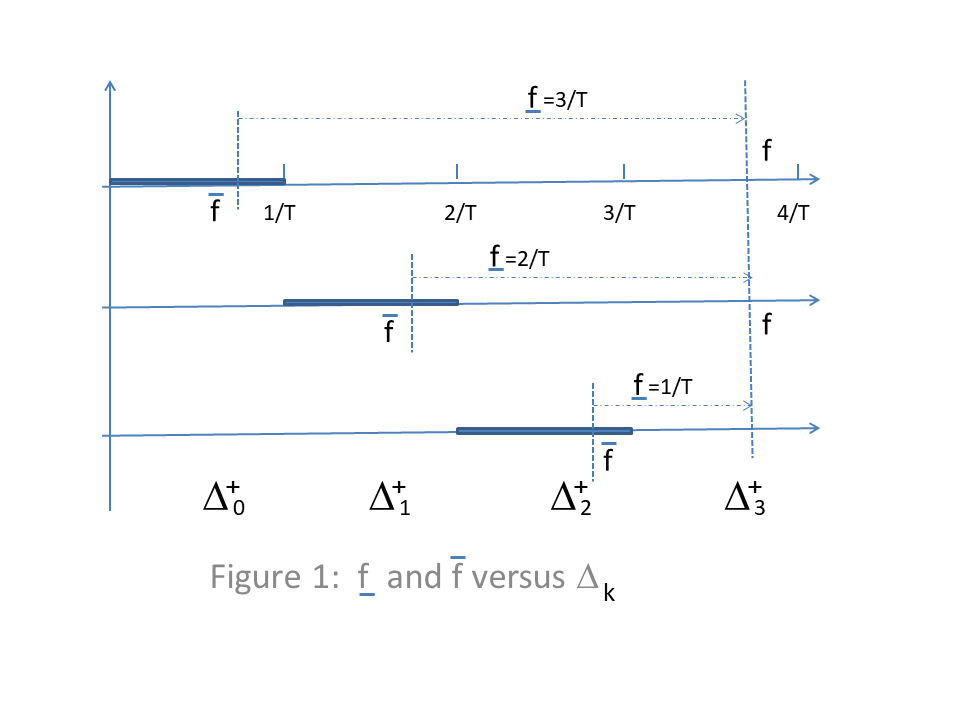}
%\caption{Tux, le pingouin}
%\label{Tux}
\end{figure}

\begin{figure}[t]
\centering
\includegraphics[width = 14cm]{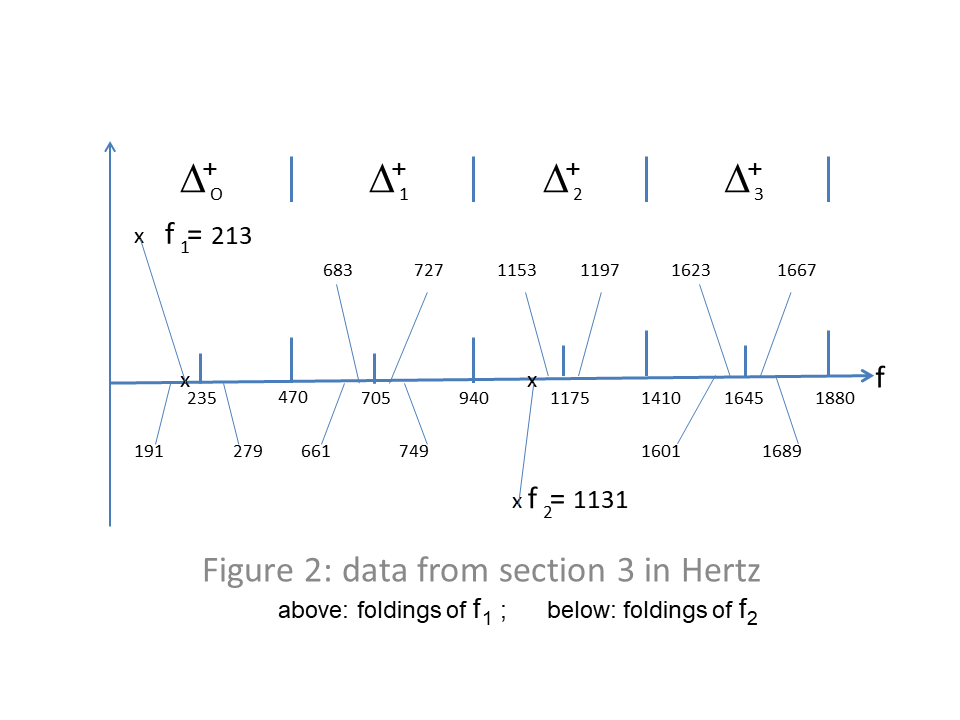}
%\caption{Tux, le pingouin}
%\label{Tux}
\end{figure}

%\FRAME{ftbpF}{7.0184in}{5.2708in}{0pt}{}{}{BTTfig1.png}{\special{language
%"Scientific Word";type "GRAPHIC";maintain-aspect-ratio TRUE;display
%"USEDEF";valid_file "F";width 7.0184in;height 5.2708in;depth
%0pt;original-width 10.4164in;original-height 7.8121in;cropleft "0";croptop
%"1";cropright "1";cropbottom "0";filename 'BTTfig1.png';file-properties
%"XNPEU";}}
%\FRAME{ftbpF}{7.0184in}{5.2708in}{0in}{}{}{BTTfig2.png}{\special%
%{language "Scientific Word";type "GRAPHIC";maintain-aspect-ratio
%TRUE;display "USEDEF";valid_file "F";width 7.0184in;height 5.2708in;depth
%0in;original-width 10.4164in;original-height 7.8121in;cropleft "0";croptop
%"1";cropright "1";cropbottom "0";filename 'BTTfig2.png';file-properties
%"XNPEU";}}

\end{document}